\documentclass[prx, preprintnumbers, amsmath, amssymb, superscriptaddress, twocolumn]{revtex4-1}
\usepackage{graphicx, color, dcolumn, natbib, bm, amssymb, amsmath,color}
\graphicspath{{fig/}}
\usepackage[final=true]{hyperref}
\usepackage[normalem]{ulem}

\begin{document}
\title{Increasing the skyrmion stability in Cu$_2$OSeO$_3$ by chemical substitution}
\author{A. S. Sukhanov}\thanks{alexandr.sukhanov@cpfs.mpg.de}
\affiliation{Max Planck Institute for Chemical Physics of Solids, D-01187 Dresden, Germany}
\affiliation{Institut f{\"u}r Festk{\"o}rper- und Materialphysik, Technische Universit{\"a}t Dresden, D-01069 Dresden, Germany}
\author{Praveen Vir}
\affiliation{Max Planck Institute for Chemical Physics of Solids, D-01187 Dresden, Germany}
\author{A. S. Cameron}
\affiliation{Institut f{\"u}r Festk{\"o}rper- und Materialphysik, Technische Universit{\"a}t Dresden, D-01069 Dresden, Germany}
\author{H. C. Wu}
\affiliation{Department of Physics, National Sun Yat-sen University, Kaohsiung 80424, Taiwan}
\author{N. Martin}
\affiliation{Laboratoire L\'eon Brillouin, CEA-CNRS, Universit\'e Paris-Saclay, CEA Saclay, 91191 Gif-sur-Yvette, France}
\author{S. M\"{u}hlbauer}
\affiliation{Heinz Maier-Leibnitz Zentrum (MLZ), Technische Universit\"{a}t M\"{u}nchen, D-85748 Garching, Germany}
\author{A. Heinemann}
\affiliation{German Engineering Materials Science Centre (GEMS) at Heinz Maier-Leibnitz Zentrum (MLZ), Helmholtz-Zentrum Geesthacht GmbH, D-85747 Garching, Germany}
\author{H. D. Yang}
\affiliation{Department of Physics, National Sun Yat-sen University, Kaohsiung 80424, Taiwan}
\author{C. Felser}
\affiliation{Max Planck Institute for Chemical Physics of Solids, D-01187 Dresden, Germany}
\author{D. S. Inosov}
\affiliation{Institut f{\"u}r Festk{\"o}rper- und Materialphysik, Technische Universit{\"a}t Dresden, D-01069 Dresden, Germany}
\date{\today}
\begin{abstract}

The cubic chiral helimagnets with the $P2_13$ space group represent a group of compounds in which the stable skyrmion-lattice state is experimentally observed. The key parameter that controls the energy landscape of such systems and determines the emergence of a topologically nontrivial magnetic structures is the Dzyaloshinskii-Moriya interaction (DMI). Chemical substitution is recognized as a convenient instrument to tune the DMI in real materials and has been successfully utilized in studies of a number of chiral magnets, such as MnSi, FeGe, MnGe, and others. In our study, we applied small-angle neutron scattering to investigate how chemical substitution influences the skyrmionic properties of an insulating helimagnet Cu$_2$OSeO$_3$ when Cu ions are replaced by either Zn or Ni. Our results demonstrate that the DMI is enhanced in the Ni-substituted compounds (Cu,Ni)$_2$OSeO$_3$, but weakened in (Cu,Zn)$_2$OSeO$_3$. The observed changes in the DMI strength are reflected in the magnitude of the spin-spiral propagation vector and the temperature stability of the skyrmion phase.

\end{abstract}

\maketitle

\section{introduction}

Since its first experimental observation in real materials~\cite{Muehlbauer,Muenzer,Pfleiderer,Yu}, the magnetic skyrmion lattice (SkL)---an ordered array of topologically protected spin whirls---is considered as a promising component of the future magnetic memory devices~\cite{Fert,Parkin}. The crucial property of any skyrmion-hosting material discovered so far is the temperature range in which the SkL can be observed. A variety of tuning parameters, such as hydrostatic~\cite{Levatic,Sidorov,Ritz,Deutsch,Wu} and negative chemical~\cite{Wu,Potapova,Dhital,Dhital2} pressure, uniaxial strain~\cite{Chacon,Nii,Fobes,Seki,Okamura}, and chemical substitution~\cite{Grigoriev,Grigoriev2,Siegfried,Grigoriev3,Bauer} were proven to be helpful in exploring the mechanism of the SkL stability in a wide class of skyrmionic materials belonging the the chiral space group (SG) $P2_13$.

In the $P2_13$ family, the generic magnetic phase diagram of a material can be pictured~\cite{Muehlbauer,Pfleiderer,Grigoriev3,Bauer}. A chiral spin helical state with a certain periodicity occurs below a Curie temperature $T_{\text C}$ as a result of competition between the ferromagnetic exchange and antisymmetric Dzialoshinskii-Moriya (DMI) interactions~\cite{Bak,Maleyev}. A weak magnetocrystalline anisotropy determines the orientation of the spiral propagation vector with respect to high-symmetry crystallographic directions. Due to degeneracy of the propagation vector orientation in the cubic system, a multi-domain helical state is formed in a macroscopic sample. Upon application of the external magnetic field, the multi-domain helical state is transformed to a single-domain state with a conical spiral propagating along the field direction~\cite{Grigoriev4,Bauer2}. In a certain range of the applied fields, the SkL is nucleated as a state that energetically competes  with the conical spiral~\cite{Muehlbauer,Roessler,Bogdanov}. A transition to the fully polarized state takes place when the Zeeman energy overcomes the energy of the DMI~\cite{Bak,Maleyev,Grigoriev4}.

The binary $B20$-type metallic and semiconducting compounds (SG $P2_13$) with the general composition $M$Si and $M$Ge (where $M = $~Mn, Fe, or Co) allows for the investigation of how the varying mixture of the 3$d$ metals affects the balance of the leading magnetic interactions in a model system. The magnetic properties in general, and stability of the SkL in particular, were extensively studied in (Fe,Co)Si by means of magnetization, ac susceptibility, specific heat, neutron spin echo spectroscopy, and small-angle neutron scattering (SANS)~\cite{Bauer3,Bannenberg,Bannenberg2,Bannenberg3}. The same set of experimental techniques were successfully applied to (Mn,Fe)Si, where a rapid suppression of the magnetic state was found when Mn is replaced by Fe in pristine MnSi~\cite{Bauer,Bannenberg4,Bannenberg5}. In the case of monogermanides, the evolution of the helical propagation vector, and the corresponding changes in the magnetic phase diagram, were tracked through the full concentration range in (Fe,Co)Ge~\cite{Grigoriev2}, (Mn,Fe)Ge~\cite{Grigoriev,Altynbaev}, and (Mn,Co)Ge~\cite{Altynbaev2}.

Unlike the monosilicides and monogermanides, which are itinerant magnets, the chiral compound Cu$_2$OSeO$_3$ (SG $P2_13$) is an insulating helimagnet~\cite{Seki2,Adams,Seki3}, where the chemical substitution of the magnetic ion can be described within the model of localized moments. Up to date, the change in magnetic structure of Cu$_2$OSeO$_3$ when Cu is replaced by the other 3$d$ elements has only been partially explored.

In a previous study~\cite{Wu2}, the substitution of Cu by nonmagnetic Zn in a series of (Cu$_{1-x}$Zn$_x$)$_2$OSeO$_3$ compounds led to an unexpected splitting of the SkL phase in the magnetic field--temperature phase diagram into two similar phase pockets separated by a small temperature interval. This observation was the first of its kind, as the SkL was only found within a single connected region of the field and temperature in all the other representatives of the $P2_13$ helimagnets. The magnetization and ac susceptibility measurements were used to track the characteristic anomalies and suggested that the splitting occurs at the substitution level $x = 0.02$ and persist up to $x = 0.15$. No such splitting of the SkL phase was observed in a later ac susceptibility study of the Ni-substituted Cu$_2$OSeO$_3$~\cite{Chandrasekhar}. 

In a recent study~\cite{Stefancic}, Stefancic \textit{et al.} found out that the observed splitting of the SkL phase in (Cu$_{1-x}$Zn$_x$)$_2$OSeO$_3$ might originate from the multiphase nature of the polycrystalline samples. According to their findings, a synthesized polycrystalline sample contains a combination of (Cu$_{1-x}$Zn$_x$)$_2$OSeO$_3$ phases with different levels of Zn substitution when a targeted Zn concentration exceeds 2\%. They were also able to demonstrate that a single crystal with $x = 0.024$ is characterized by only one SkL phase. As was shown in Ref.~\cite{Wu2}, Zn occupies only the Cu(II) site in the polycrystalline samples, which determines the properties of the substituted compounds. However, it remains unclear if the same site is occupied by Zn in the single crystals studied in~\cite{Stefancic}. Despite the fact that the splitting of the SkL phase is being debated, the influence of chemical substitution on the energy landscape of the fundamental magnetic interactions in Cu$_2$OSeO$_3$ remains an open and interesting question.

In this paper, we utilize SANS to experimentally observe how the magnetic phase diagram and the helimagnetic propagation vector change in the insulating chiral magnet Cu$_2$OSeO$_3$ when the chemical substitution of the magnetic ion is applied. We systematically studied a series of polycrystalline (Cu,Zn)$_2$OSeO$_3$ and (Cu,Ni)$_2$OSeO$_3$ samples. Particularly, in our study we focused on the substitution-dependent change of the magnetic propagation vector and its relation to the enhancement of the temperature stability of the SkL.

\section{experimental details}

\begin{table}[t]
\small\addtolength{\tabcolsep}{+7pt}
\caption{The substitution levels of Zn and Ni in the series of (Cu$_{1-x}$Ni$_x$)$_2$OSeO$_3$ and (Cu$_{1-y}$Zn$_y$)$_2$OSeO$_3$ samples used in the present study. The Curie temperatures $T_{\text C}^{(1)}$ and $T_{\text C}^{(2)}$ refer to magnetic transitions of two phases with different concentration of Zn or Ni, where applicable.}
\label{tab:tab1}
 \begin{tabular}{c c c}
 \hline\hline
 Nominal & Actual~\cite{Supp} & Curie temperature \\ [1ex] 
 \hline
 $x = 0$ & $x = 0$ & $T_{\text C} = 57.7$ K\\ [1ex] 
 \hline
 $x = 0.04$ & $x = 0.04$ & $T_{\text C} = 54.6$ K\\ [1ex] 
 \hline
 $x = 0.08$ & $x_1 = 0.021$ & $T_{\text C}^{(1)} = 56.1$ K\\ [1ex] 
    & $x_2 = 0.063$ & $T_{\text C}^{(2)} = 52.8$ K\\ [1ex]
    \hline
 $y = 0.04$ & $y_1 = 0.008$ & $T_{\text C}^{(1)} = 55.8$ K\\ [1ex]
    & $y_2 = 0.025$ & $T_{\text C}^{(2)} = 51.6$ K\\ [1ex]
    \hline
 $y = 0.10$ & $y_1 = 0.011$ & $T_{\text C}^{(1)} = 55.0$ K\\ [1ex]
    & $y_2 = 0.055$ & $T_{\text C}^{(2)} = 44.5$ K\\ [1ex]  
 
 \hline\hline
\end{tabular}
\end{table}

Polycrystalline samples with the nominal composition $M_2$OSeO$_3$, where $M = $ Cu, Cu$_{0.96}$Zn$_{0.04}$, Cu$_{0.9}$Zn$_{0.1}$, Cu$_{0.96}$Ni$_{0.04}$, and Cu$_{0.92}$Ni$_{0.08}$, were synthesized by a solid-state reaction as described elsewhere~\cite{Wu2,Chandrasekhar,Stefancic}. The magnetic characterization of the synthesized samples revealed a coexistence of two phases with different Zn content in the samples with nominal compositions (Cu$_{0.96}$Zn$_{0.04}$)$_2$OSeO$_3$ and (Cu$_{0.9}$Zn$_{0.1}$)$_2$OSeO$_3$ in full agreement with Ref.~\cite{Stefancic}. In a close analogy with the pair of Zn-substituted compounds, a phase separation occurred in the sample with a high Ni concentration (supplementary figure Fig.~S2~\cite{Supp}). Moreover, a rigorous analysis of the XRD patterns (supplementary note~1) revealed that the actual average composition of the sample with the nominal 8\%~Ni content differs significantly and amounts only to $\approx$~2.5\% Ni substitution due to a presence of an impurity. The information on the targeted and actual substitution levels along with the corresponding magnetic ordering temperatures $T_{\text C}$, inferred from the dc susceptibility measurements, are summarized in Table~\ref{tab:tab1}.

SANS experiments were conducted at the instruments SANS-1 (FRM-II, Garching, Germany) and PA20 (LLB-Orph\'{e}e, CEA Saclay, France)~\cite{PA20}. For a measurement of each compound, $\sim 0.5$~g of the powder sample was filled into an aluminium cylindrical sample holder. The samples with the Ni-2.1\% and Zn-1.1\% substitution levels were measured at SANS-1, measurements of the Ni-4\% and Zn-0.8\% samples were carried out at PA20. In addition, the pure Cu$_2$OSeO$_3$, as the reference compound, was measured on both instruments to eliminate  potential errors related to a sample temperature offset or resolution effects. An accurate comparison of the results of magnetic characterization and SANS measurements of the reference compound (presented by the exact same sample) allowed us to detect a small constant temperature offset present during the PA20 measurements, which was corrected during the data analysis.

To access a relevant momentum transfer range, the incident neutron wavelength $\lambda$ of 6~\AA ~($\Delta\lambda$/$\lambda = 10$\%) and detector distance of 20~m were set in the SANS-1 experiment. In the PA20 measurements, the configurations with detector distances of 18.78 and 15~m and $\lambda = 6$~\AA ~($\Delta\lambda$/$\lambda = 11.7$\%) was used. A sample temperature stabilization not worse than $\pm 50$~mK was kept during all the measurements. The same field protocol was applied to collect the SANS data of every compound: the sample was zero field cooled to a temperature below $T_{\text C}$, afterwards the magnetic field was applied. The SANS patterns were collected in a magnetic field, which was increased stepwise at a constant temperature. The field was then driven back to zero value, followed by a zero field change of the sample temperature, after which the field-scan measurements were repeated.

\section{results}

\subsection{SkL in Cu$_2$OSeO$_3$}\label{sec:III.A}

\begin{figure*}
\includegraphics[width=\linewidth]{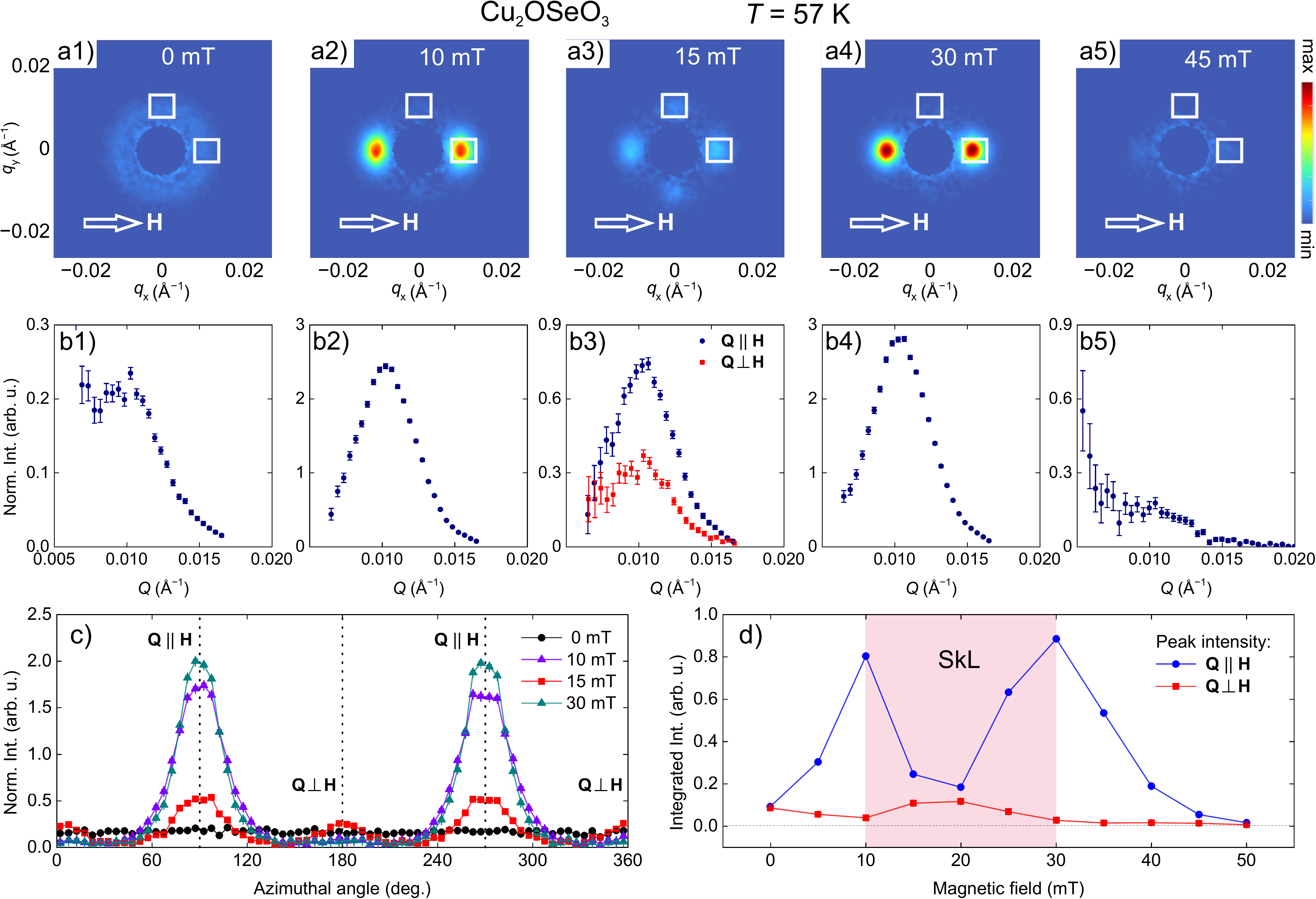}\vspace{3pt}
        \caption{(color online). The SkL in polycrystalline Cu$_2$OSeO$_3$ at 57~K. (a1)--(a5) SANS patterns collected at the SANS-1 instrument. Different values of the applied magnetic field are indicated at the detector images. The magnetic field is applied perpendicular to the neutron beam. (b1)--(b5) The corresponding normalized SANS intensity as a function of the momentum transfer along and perpendicular to the field direction. (c) The normalized azimuthal intensity profiles at different magnetic fields extracted from the SANS patterns. The radial integration range of (0.005--0.015)~\AA$^{-1}$ was applied. The solid lines are a guide to the eye. (d) The integral intensity of the Bragg peaks as a function of the field. The integration areas are shown by the white boxes in (a1)--(a5). The solid lines are a guide to the eye. The SkL phase is highlighted by the shaded area. A measurement at 65~K ($T>T_\text{C}$) was used for a background subtraction for all the data shown.}
        \label{ris:fig1}
\end{figure*}

Before discussing the results obtained in measurements of the substituted compounds, a typical SANS response of a polycrystalline Cu$_2$OSeO$_3$ sample has to be examined in detail. In what follows, we consider the intensity of SANS, collected from the sample of Cu$_2$OSeO$_3$ at a constant temperature of 57~K, as a function of momentum transfer and applied magnetic field. According to previous studies of Cu$_2$OSeO$_3$ single crystals~\cite{Seki2,Adams,Seki3,Qian,Levatic,Bannenberg6}, 57~K lies within the temperature range at which the SkL can be stabilized by an applied magnetic field of 10 to 30~mT.

Figures~\ref{ris:fig1}(a1)--\ref{ris:fig1}(a5) demonstrate SANS patterns collected from the sample in the different magnetic states: multidomain helical, monodomain conical at $H \simeq H_{\text a1}$, SkL, monodomain conical at $H \simeq H_{\text a2}$, and field-polarized, respectively. The pattern at zero field [Fig.~\ref{ris:fig1}(a1)] is characterized by a uniform ring of intensity formed by the helical structure of randomly oriented crystallites of the powder sample. The ring of intensity can be imagined as a sphere of intensity in 3D reciprocal space crossed by the detector plane. In an applied field of 10~mT, the helical domains of every crystallite are rotated along the field direction and form two intense spots of intensity, as shown in Fig.~\ref{ris:fig1}(a2). When the field is increased to 15~mT, the intensity along the field drops drastically, whereas it appears at the same $\lvert \textbf{Q} \rvert$ in the perpendicular direction [Fig.~\ref{ris:fig1}(a3)]. The latter is a characteristic scattering from the SkL, which is seen in SANS as a diffraction pattern in the plane normal to the field. The SkL in each crystallite creates the six-spot diffraction pattern~\cite{Adams,Seki3}. Such a pattern is weakly pinned to certain in-plane (with respect to the plane singled out by the field) crystallographic directions defined by a weak magnetocrystalline anisotropy~\cite{Seki3,Bannenberg6}. Due to powder averaging, the total scattering from the SkL gives rise to a ring of intensity that forms two vertical spots of intensity upon intersection with the detector plane. At a higher field of 30~mT, the monodomain conical phase is restored, as evidenced in Fig.~\ref{ris:fig1}(a4) by the absence of intensity away from the field direction. No intensity can be seen in a pattern collected in the field-polarized state [Fig.~\ref{ris:fig1}(a5)].

The corresponding profiles of SANS intensity are demonstrated in Figs.~\ref{ris:fig1}(b1)--\ref{ris:fig1}(b5). The intensity distribution from SANS maps presented in Figs.~\ref{ris:fig1}(a1)--\ref{ris:fig1}(a5) was plotted as a function of momentum that is either parallel or perpendicular to the field direction, with the latter applicable only for Fig.~\ref{ris:fig1}(b3). The profile shown in Fig.~\ref{ris:fig1}(b1) is characterized by a weak Bragg peak from the powder-averaged spin-spiral texture. The peak position is observed at $\sim 0.01$~\AA$^{-1}$, which is in agreement with previous single crystal studies~\cite{Adams,Seki3,Bannenberg6}. The peak intensity increases by $\sim 10$ times in the conical state, where the entire helimagnetic texture in the sample is oriented along the field [Figs.~\ref{ris:fig1}(b2), \ref{ris:fig1}(b4)]. Figure~\ref{ris:fig1}(b3) compares the intensity profiles in two orthogonal directions [$q_x$ and $q_y$ in Fig.~\ref{ris:fig1}(a3)] in the SkL state. Although most of the sample is occupied by the skyrmion phase at 15~mT, the Bragg intensity of the SkL is weaker due to the fact that it is distributed in reciprocal space across the ring that is perpendicular to the detector plane. At the same time, the remaining conical phase, being in significantly smaller fraction, fully contributes its scattering intensity into the in-detector-plane Bragg reflection.

The presence of the SkL in the polycrystalline sample of Cu$_2$OSeO$_3$ can be also identified by considering the azimuthal profiles (the angle in the detector plane) of SANS intensity, as shown in Fig.~\ref{ris:fig1}(c). The scattering from the multidomain helical state at $B = 0$~mT is characterized by the intensity that is independent of the azimuthal angle. The azimuthal symmetry is broken at $B = 10$~mT, when two strong Bragg peaks with FWHM of $\sim 35^{\circ}$ are seen at positions of 90$^{\circ}$ and 270$^{\circ}$ ($\textbf{Q} \parallel \textbf{H}$). In addition to this, two extra Bragg peaks with the same FWHM appear in 15~mT at 0$^{\circ}$ and 180$^{\circ}$ ($\textbf{Q} \perp \textbf{H}$).

In order to precisely determine the phase boundaries of the SkL: $H_{\text a1}$ and $H_{\text a2}$, the intensity in each collected SANS pattern was integrated over the boxes enclosing the Bragg peaks for $\textbf{Q} \parallel \textbf{H}$ and $\textbf{Q} \perp \textbf{H}$, and plotted as a function of magnetic field in Fig.~\ref{ris:fig1}(d). The integral intensity for $\textbf{Q} \parallel \textbf{H}$ increases drastically from 0 to 10~mT, where it reaches its maximum followed by a pronounced dip with the minimum at 20~mT. Correspondingly, the integral intensity of the  $\textbf{Q} \perp \textbf{H}$ Bragg peak demonstrates a maximum at $B = 20$~mT. At a field of 30~mT, the intensity of the SkL phase vanishes and the intensity of the conical phase increases back to the value that is approximately equal to the conical intensity at 10~mT. Above $B = 30$~mT, the intensity of the SkL is absent and the scattering from the conical helix monotonically decreases until it disappears completely at $\sim 45\text{--}50$~mT. The characteristic dip in the $I(H)$ curve (for $\textbf{Q} \parallel \textbf{H}$) is specific for high temperatures at which the SkL phase exists and cannot be observed at low temperatures (see the supplementary Fig.~S3 for an example of $I(H)$ at $T = 5$~K~\cite{Supp}).

\subsection{Helimagnetic structure in (Cu,Ni)$_2$OSeO$_3$ and (Cu,Zn)$_2$OSeO$_3$}

\begin{figure}[t]
        \begin{minipage}{0.85\linewidth}
        \center{\includegraphics[width=1\linewidth]{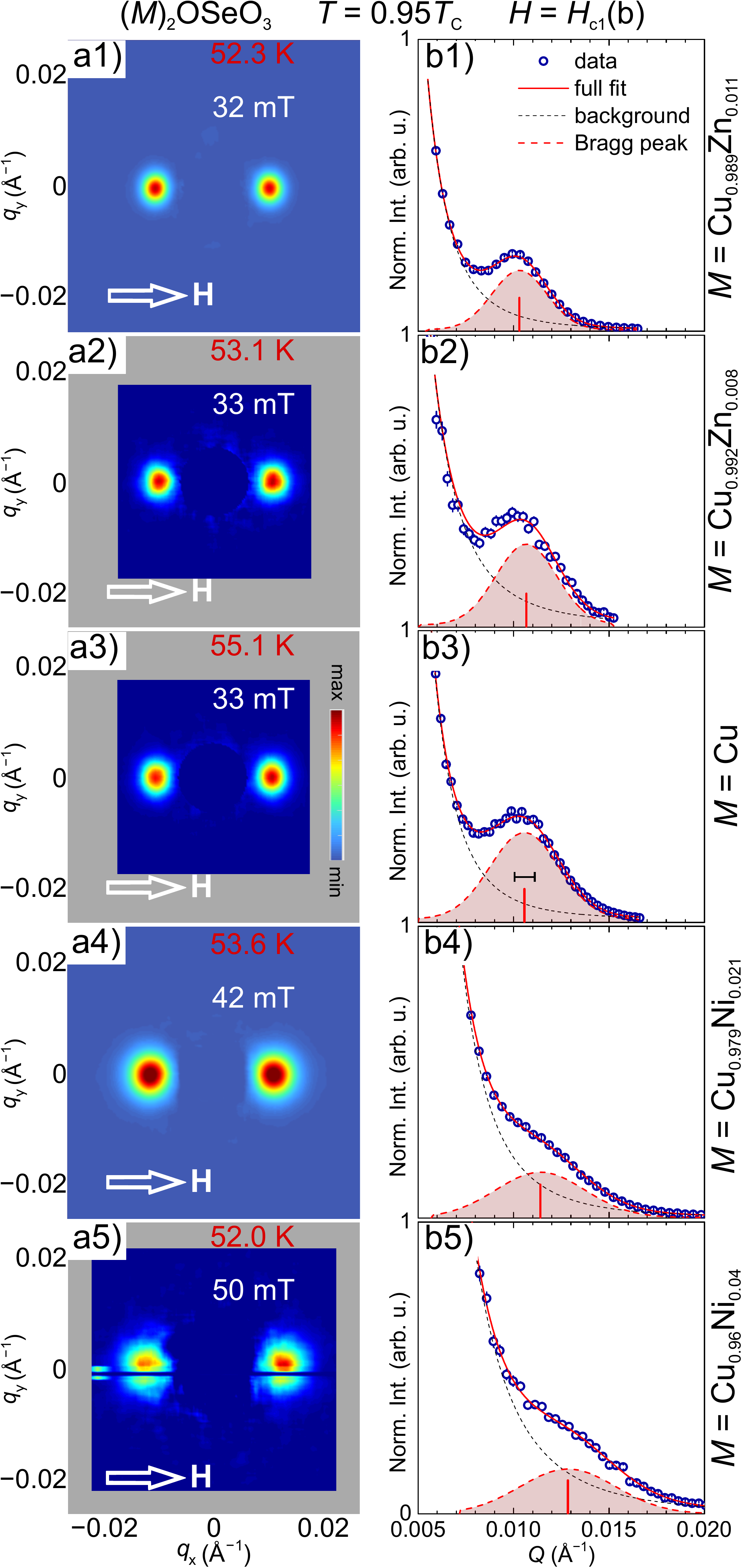}}
        \end{minipage}
        \caption{(color online). The helimagnetic structure of (Cu,Ni)$_2$OSeO$_3$ and (Cu,Zn)$_2$OSeO$_3$. (a1)--(a5) SANS patterns collected from the compounds with different substitution levels of Zn and Ni. The detector images in (a1) and (a4) were taken in measurements on the SANS-1 instrument, (a2),(a3), and (a5) represent data from PA20. The masked horizontal line in (a5) is due to an inactive detector tube. The background subtraction was applied to all the diffraction patterns. (b1)--(b5) The corresponding intensity profiles without the background subtraction. The solid and dashed lines show the results of the fit as described in the legend of (b1). The vertical bars show the extracted Bragg positions.}
        \label{ris:fig2}
\end{figure}

\begin{figure}[t]
        \begin{minipage}{0.99\linewidth}
        \center{\includegraphics[width=1\linewidth]{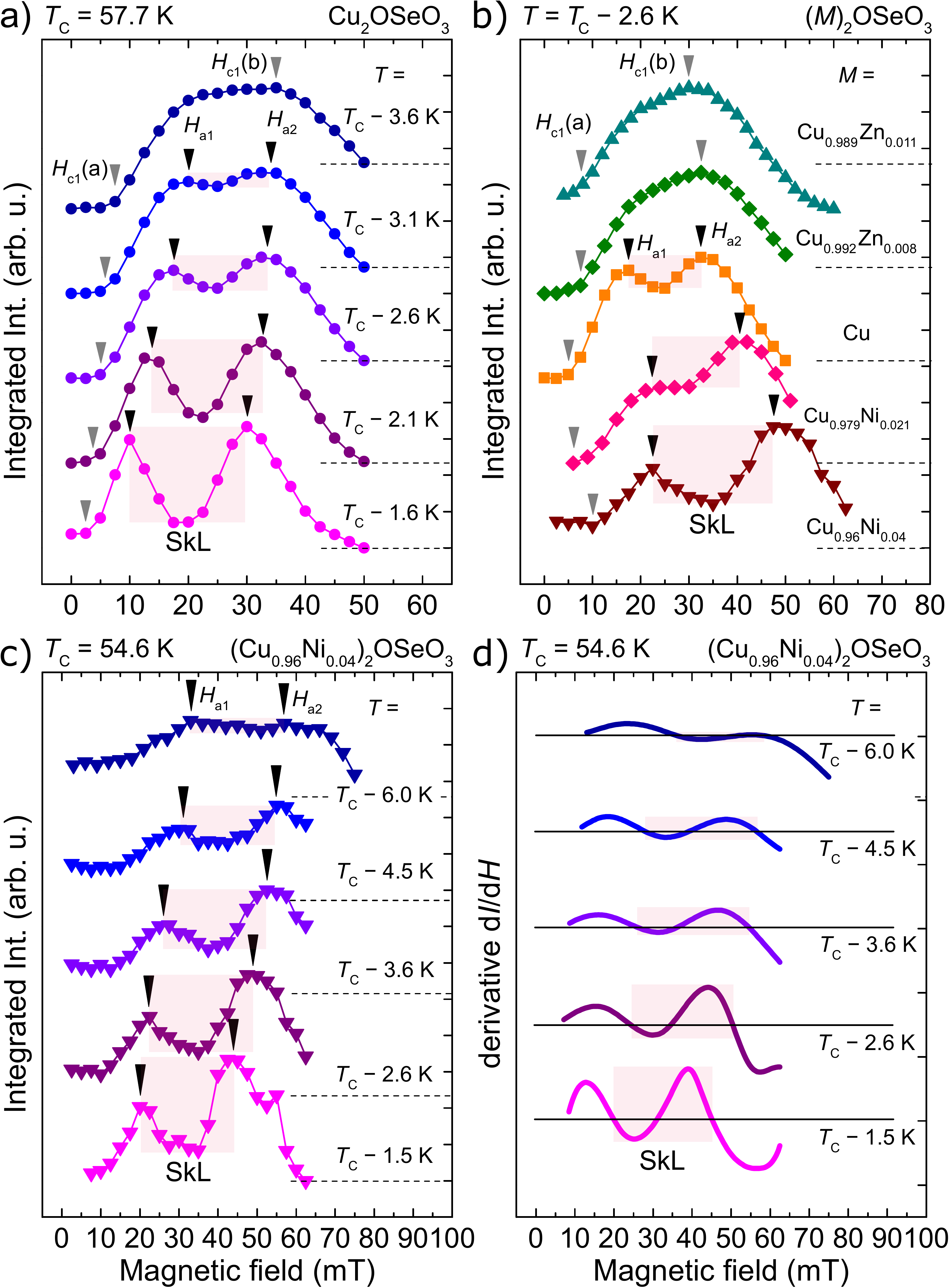}}
        \end{minipage}
        \caption{(color online). Integral intensity of the helical Bragg peak $I$ as a function of magnetic field at different temperatures just below $T_{\text C}$ in (Cu,Ni)$_2$OSeO$_3$ and (Cu,Zn)$_2$OSeO$_3$. (a) The $I(H)$ curves showing the SkL phase observed in the pure Cu$_2$OSeO$_3$ compound. The black arrows denote the boundaries of the SkL phase $H_{\text a1}$ and $H_{\text a2}$. The gray arrows point to the critical fields $H_{\text c1}$ of the transition from the multidomain helical to the monodomain conical phase. (b) $I(H)$ at $T = T_{\text C} - 2.6$~K measured in samples with different substitution. (c) The $I(H)$ curves of the Ni-4\% doped compound. The data were offset for a clarity as indicated by the dashed lines. The error bars are smaller than the symbol size, the solid lines are a guide to the eye. (d) The first derivative of the data presented in (c). The SkL phase is highlighted by the shaded area. The calculated derivative at each temperature was smoothed by a cubic B-spline function with the same smoothing factor for clarity.}
        \label{ris:fig3}
\end{figure}

\begin{figure*}
\includegraphics[width=\linewidth]{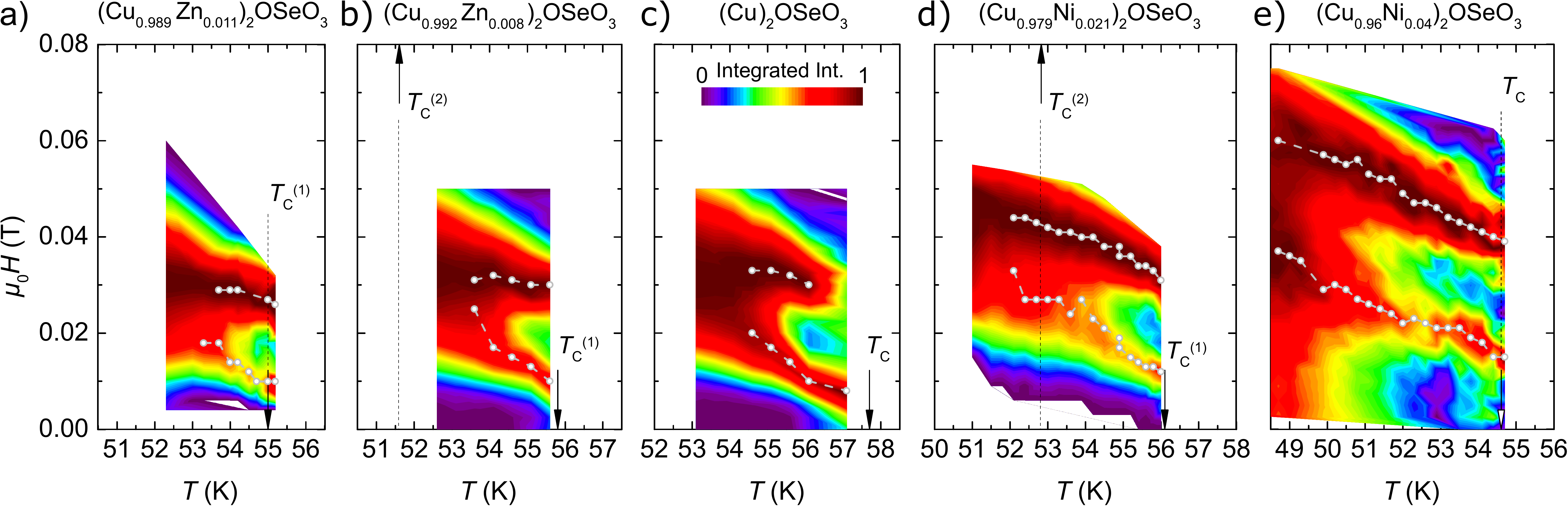}\vspace{3pt}
        \caption{(color online). The magnetic phase diagrams of Cu$_2$OSeO$_3$ with different chemical substitution extracted from SANS data. The corresponding chemical formulas of the studied compounds are shown in (a)--(e). The labels $T_{\text C}^{(2)}$ in (b) and (d) refer to the transition temperature of the secondary phase as explained in the text. The white circles denote the phase boundaries of the SkL state inferred from the $I(H)$ curves as discussed in the text.}
        \label{ris:fig4}
\end{figure*}

Having discussed the typical SANS patterns that were obtained in measurements of a polycrystalline Cu$_2$OSeO$_3$, we turn to the question of the magnetic structure of the substituted compounds. To directly compare the scattering from a spin texture in the Zn- and Ni-substituted materials, Figs.~\ref{ris:fig2}(a1)--\ref{ris:fig2}(a5) show the SANS patterns  collected from each compound at the same reduced temperature $T/T_{\text{C}}$ and in magnetic fields that correspond to the helical-to-conical (or the SkL-conical, if relevant) metamagnetic transition $H_{\text c1}$(b) (or $H_{\text a2}$). The data obtained from each compound is shown at the same conditions $T = 0.95T_{\text C}$ and $H = H_{\text c1}$ for a direct comparison of the observed propagation vector. The critical fields were determined by measuring the integral intensity at the Bragg position along the field direction as a function of the field magnitude. Such an approach is similar to the analysis presented in Fig.~\ref{ris:fig1}(d). As can be seen, each substituted compound demonstrates a Bragg peak that is similar to the one resolved in the parent compound. This observation confirms that all the substituted compounds are also helimagnets. The temperature-dependent measurements of the helical peak intensity showed good agreement with the Curie temperatures determined from the bulk magnetic measurements (supplementary Fig.~S4~\cite{Supp}). The absolute values of the field that satisfies the chosen conditions $H = H_{\text c1}$(b) at $T = 0.95T_{\text{C}}$ stays unchanged at $\approx 32$--33 mT in the Zn-substitution series. In contrast, the samples with the Ni substitution are characterized by the critical fields increasing to $\approx 42$~mT in the Ni-2.1\% compound and to $\approx 50$~mT in the Ni-4\% one, which is more than 1.5~times higher.

To determine the spin-spiral period, the SANS intensity was plotted as a function of the momentum transfer, as drawn in Figs.~\ref{ris:fig2}(b1)--\ref{ris:fig2}(b5). A fit by a Gaussian function was applied to the Bragg peak  in each intensity profile. In a close analogy with $H_{\text c1}$(b), the magnitude of the propagation vector $k_{\text{s}}$ shows only a small change between the parent compound and the samples with the Zn-0.8\% and Zn-1.1\% substitution level, whereas it becomes significantly larger when the substitution by Ni is utilized: $k_{\text{s}}$ (Ni-4\%) = 0.01285(8)~\AA$^{-1}$, which corresponds to the real-space period of 489~\AA.

The instrumental resolution, which is $\sim 10$\%, is indicated in the $I(Q)$ profile of Fig.~\ref{ris:fig2}(b3). As can be seen, the Bragg peaks exhibit a width that is broader than the experimental resolution. This occurs due to the multiple Porod scattering, which is frequently observed in SANS on polycrystalline samples. Porod's law describes the asymptotic scattering from the surfaces of inhomogeneities (the crystalline grains in the present case) and has the form $I(Q) \sim Q^{-4}$, valid for $Q > 1/L$, where $L$ is the size of crystallites. The influence of the Porod scattering on the $I(Q)$ profiles shown in Figs.~\ref{ris:fig2}(b1)--\ref{ris:fig2}(b5) is twofold. First, it results in an intense scattering for $Q < k_{\text s}$. Second, the neutrons scattered at the Bragg angle, undergo subsequent multiple Porod scattering that leads to the observed broadening of the peak (due to the finite sample thickness). The fact that the pure compound demonstrates the same broadening suggests that the broadening of the Bragg peaks of the substituted compounds is not caused by the chemical substitution.

\subsection{Changes in stability of the SkL phase}

The temperature stability of the SkL phase can be explored by measuring the intensity of the helical Bragg peak as a function of magnetic field at different constant temperatures after the ZFC procedure. As has been discussed in Sec.~\ref{sec:III.A}, the onset of the SkL nucleation is manifested by suppression of the scattered intensity from the conical phase. The $I$ vs $H$ curves obtained in measurements of Cu$_2$OSeO$_3$ as the reference compound are shown in Fig.~\ref{ris:fig3}(a). The data at $T = T_{\text C} - 1.6$~K correspond to a temperature at which the SkL is stable for any orientation of the field with respect to the crystallographic directions~\cite{Adams}. This fact is supported by the full dip at $\approx 20$~mT in the $I(H)$ data, which implies that the entire volume of a polycrystalline sample is occupied by the SkL phase. At a lower temperature of $T_{\text C} - 2.1$~K, the depth of the minimum in $I(H)$ decreases. This shows that only a half of the randomly oriented crystallites satisfy the condition of the mutual orientation of the applied field and the crystallographic axes for the SkL nucleation at 55.6~K. The dip becomes shallow at lower temperatures and can be distinguished down to $T = T_{\text C} - 3.1$~K.

In order to directly compare the temperature range of the SkL stability in the series of substituted compounds, $I(H)$ data of every compound were plotted at a temperature 2.6~K lower than the $T_{\text C}$ of each compound. As can be seen in Fig.~\ref{ris:fig3}(b), the substitution of Cu by Zn leads to the suppression of the SkL. This implies that the SkL pocket contracts towards $T_{\text C}$. On the contrary, the compounds with Ni substitution demonstrate the enhancement of the SkL. The depth of the intensity drop increases by a factor of $\sim 1.6$ in the Ni-2.1\% sample. Moreover, the sample with the Ni-4\% content is characterized by an almost full dip in $I(H)$, which was seen in the reference compound only at elevated temperatures. Importantly, the chemical substitution leads to an approximately equal rise of the critical field $H_{\text c1}$(a) regardless whether Cu was replaced by Zn or Ni. This agrees with expectations based on the definition of $H_{\text c1}$(a): the chemical disorder enlarges a concentration of the pinning centers in the system. The latter is proportional to the impurity concentration and independent from the impurity type.

In contrast to $H_{\text c1}$(a), the phase boundaries of the SkL, $H_{\text a1}$ and $H_{\text a2}$, reflect the change in the energy landscape of the system inherent to a specific substituted element. As can be seen in Fig.~\ref{ris:fig3}(b), the enhancement of the SkL stability is accompanied by a proportional increase of both $H_{\text a1}$ and $H_{\text a2}$ in the Ni-substituted compounds. The $T$-range at which the SkL can be stabilized in magnetic field in (Cu$_{0.96}$Ni$_{0.04}$)$_2$OSeO$_3$ can be directly compared to that of the reference (pure) compound as depicted in Figs.~\ref{ris:fig3}(c) and \ref{ris:fig3}(a). The $I(H)$ curve at $T = T_{\text C} - 1.5$~K of the Ni-4\% sample [Fig.~\ref{ris:fig3}(c)] qualitatively reproduces the $I(H)$ curve of the pure compound at the same relative temperature [Fig.~\ref{ris:fig3}(a)]. The SkL is found within a 20~mT wide field region in both compounds but shifted up by $\sim 10$~mT in the case of the Ni-4\% compound. Similarly to the pure compound, the Ni-4\% material demonstrates a gradual reduction in the intensity drop associated with the SkL formation at lower temperatures. However, the SkL in Cu$_2$OSeO$_3$ is already absent at $T = T_{\text C} - 3.6$~K, whereas a clear dip in $I(H)$ is still observed in the Ni-substituted compound. The field-dependent change of the scattered intensity at various temperatures can also be analyzed by its first derivative d$I$/d$H$ [Fig.~\ref{ris:fig3}(d)]. As the temperature decreases, d$I$/d$H$ flattens out. The critical fields $H_{\text a1}$ and $H_{\text a2}$ can be determined by the first and  third roots of the equation d$I$/d$H = 0$.

\subsection{Full magnetic phase diagrams mapped by SANS}

The influence of the doping by different 3\textit{d} elements on the magnetic structure of Cu$_2$OSeO$_3$ can be illustrated when the phase diagrams of all the samples are compared simultaneously. An example analysis of $I(H)$ (the field dependence of the integral intensity of the helical Bragg peak) was  demonstrated in Figs.~\ref{ris:fig3}(a)--\ref{ris:fig3}(c). The same approach can be used to determine the precise boundary of the SkL in the $(T$--$H)$ parameter space for every compound discussed in the present study. The $I(H)$ curves, obtained in each field-sweep measurement at a constant temperature, were combined in a single $I(H,T)$ set and plotted as a colormap in Figs.~\ref{ris:fig4}(a)--\ref{ris:fig4}(e).

In the vicinity of $T_{\text C}$, the magnetic phase diagram of the pure compound [Fig.~\ref{ris:fig4}(c)] is characterized by the SkL that is stabilized in a field of 8~mT at $T = 57.1$~K ($T_{\text C} - 0.6$~K) and persists up to $\sim 30$~mT, where a smeared transition to the field-polarized state takes place. The SkL phase keeps the same field width of 20~mT at $T = 56.1$~K, after which the field region of the SkL contracts down to a width of 13~mT at 54.6~K. Below $T = 54.6$~K, the SkL pocket closes rapidly, and no traces of the SkL presence can be observed at lower temperatures. Overall, the SkL in the reference compound is found within a temperature region of 3.1~K below $T_{\text C}$. As evidenced by Fig.~\ref{ris:fig4}(c), the critical fields $H_{\text a1}$ and $H_{\text a2}$ exhibit the following temperature change: $H_{\text a2}$ keeps approximately the same value (changes by less than 10\%), whilst $H_{\text a1}$ increases by a factor of $\sim 2.5$ from high to low temperature.

The phase-boundary loop enclosing the SkL phase in the $(H$--$T)$ diagram of the Zn-substituted samples [Figs.~\ref{ris:fig4}(a)--\ref{ris:fig4}(b)] can be described by similar features: a $T$-independent $H_{\text a2}$ line and a much more pronounced change in the magnitude of $H_{\text a1}(T)$. The critical field $H_{\text a1}$ was found to take values from 10~mT ($T \simeq T_{\text C}$) to 25~mT ($T = T_{\text C} - 2.2$~K) in the Zn-0.8\% compound, and from 10 to 18~mT in the compound with Zn-1.1\% substitution level; $H_{\text a2}$ amounts to $\sim 31$~mT and $\sim 28$~mT for the Zn concentration of 0.8\% and 1.1\%, respectively. The $T$-range at which the SkL exists shows a decreasing trend upon the Zn substitution: from 3.1~K below $T_{\text C}$ in the pure compound to 1.7~K-wide $T$ region in (Cu$_{0.989}$Zn$_{0.011}$)$_2$OSeO$_3$.

A different situation occurs when Ni substitution is applied, as shown by the phase diagrams of the Ni-2.1\% and Ni-4\% samples in Figs.~\ref{ris:fig4}(d)--\ref{ris:fig4}(e). First, the temperature stability of the SkL is noticeably enhanced. Second, $H_{\text a2}$ starts showing ф temperature dependence with a linear slope similarly to $H_{\text a1}$. Furthermore, the critical fields proportionally shift to higher values. For example, $H_{\text a1}(T_{\text C}) = 15$~mT and $H_{\text a2}(T_{\text C}) = 39$~mT in the Ni-4\% sample. Also, in the vicinity of the low-temperature boundary of the SkL, $H_{\text a2}$ reaches a value of 60~mT, which is two times higher than that in the reference compound.

\section{discussion}

\begin{figure}[t]
        \begin{minipage}{0.99\linewidth}
        \center{\includegraphics[width=1\linewidth]{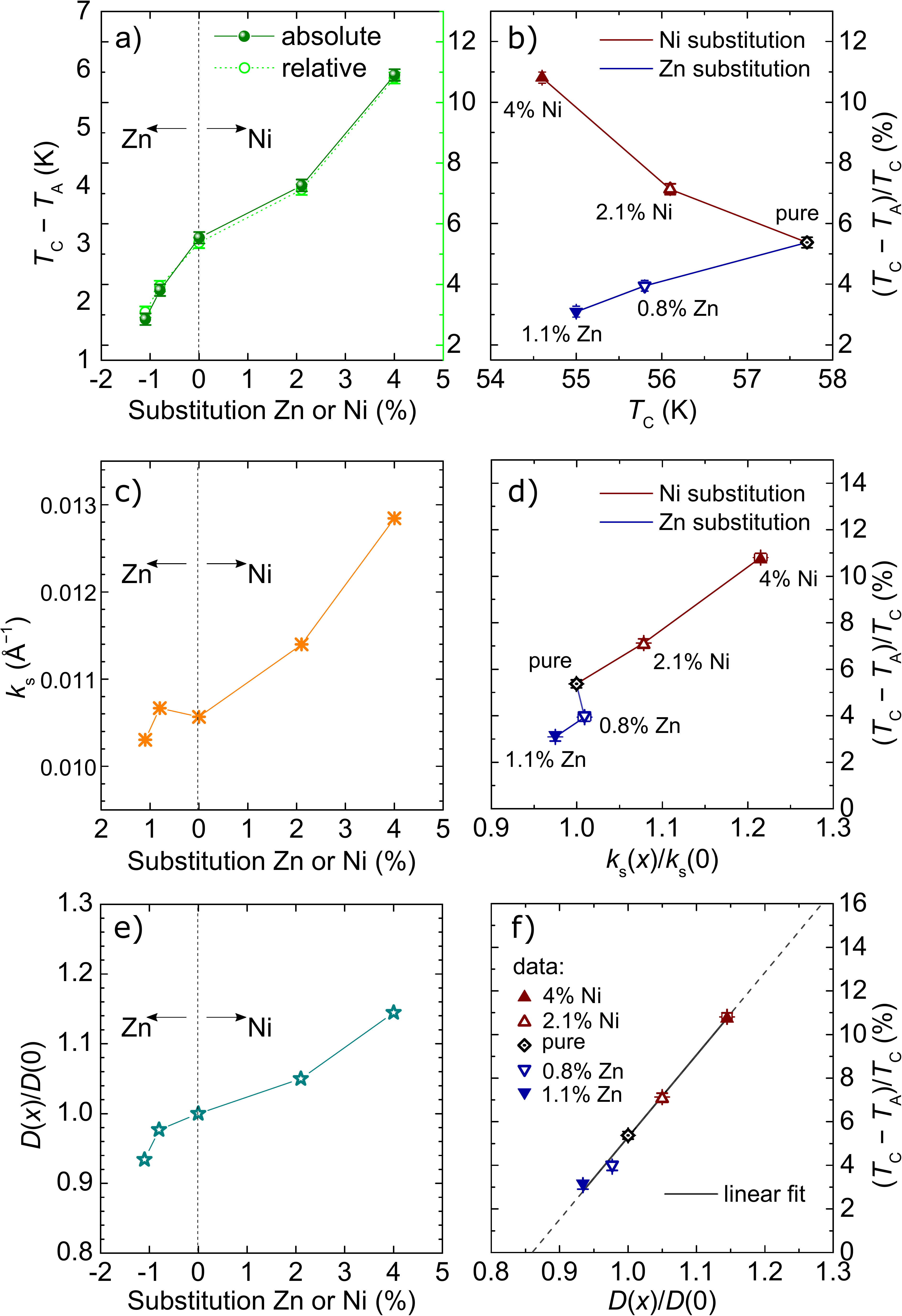}}
        \end{minipage}
        \caption{(color online). The correlation between the experimentally extracted microscopic parameters. (a) The temperature range of the SkL phase in the substituted compounds, $T_{\text A}$ is the lowest temperature at which the SkL can be nucleated in the applied magnetic field (absolute in K and normalized to $T_{\text C}$). (b) The SkL stability versus $T_{\text C}$ for all the studied compounds. (c) The helical propagation vector of each compound. (d) The SkL temperature stability versus the normalized propagation vector, $k_{\text s}(0)$ is the propagation vector of pure Cu$_2$OSeO$_3$. (e) The extracted values of the DMI constant for samples with different substitution level. The solid lines are a guide to the eye. (f) The SkL stability versus the normalized DMI, $D(0)$ stands for the DMI strength in pure Cu$_2$OSeO$_3$. The solid line is a linear fit, the dashed line is a linear extrapolation.}
        \label{ris:fig5}
\end{figure}

As evidenced by our results, the replacement of Cu ions by nonmagnetic Zn atoms causes changes in the magnetic structure and magnetic phase diagram of the compound. On the mean-field level, $T_{\text C}$ is proportional to the exchange interaction between a pair of spins and the coordination number, which is reduced when some of the spins are replaced by voids. The nonmagnetic dilution reduces the average number of Cu-Cu bonds, which leads to a reduction of $T_{\text C}$. In the limit of a small impurity concentration, the corresponding reduction of $T_{\text C}$ is expected to be linear. The change in the propagation vector upon the substitution does not exceed 5\%, whereas the SkL phase shows a clear tendency to decrease its temperature stability range, which amounts to $\sim 30$\% (by $\sim 1$~K) shrink over $x = 0.011$.

Figures \ref{ris:fig5}(a)--\ref{ris:fig5}(f) summarize the change in the key parameters of the studied system. The absolute temperature width of SkL phase pocket, defined as $T_{\text C} - T_{\text A}$ (where $T_{\text A}$ is the lowest temperature at which the SkL is observed), and the relative quantity $\left(T_{\text C} - T_{\text A}\right)/T_{\text C}$ (normalized to the $T_{\text C}$), are shown in Fig.~\ref{ris:fig5}(a) as a function of the concentration levels of both Zn and Ni substituents. As can be seen, Ni and Zn cause an opposite influence on the SkL: whilst substitution by Zn suppresses the phase, Ni substitution improves its temperature stability. Moreover, Ni leads to a much steeper change.

Importantly, the temperature stability of the SkL phase is not related to the trend that $T_{\text C}(x)$ demonstrates. Figure \ref{ris:fig5}(b) shows the relative SkL phase width $\left(T_{\text C} - T_{\text A}\right)/T_{\text C}$ of a compounds versus the $T_{\text C}$ of the compound. Both Zn- and Ni-substituted compounds exhibit decrease in $T_{\text C}$, but the impact on the SkL is clearly opposite for the two series.

The shift in the balance of magnetic interactions is evidenced in Fig.~\ref{ris:fig5}(c), where the magnitude of the spiral propagation vector is drawn as a function of Zn or Ni concentration. The propagation vector is barely affected when Zn is placed, whereas $k_{\text s}$ increases by 20\% in the compound with $x (\text{Ni}) = 0.04$. Notably, the trend in $k_{\text s} = k_{\text s}(x)$ closely resembles the concentration dependence of the SkL stability. When the quantity $\left(T_{\text C} - T_{\text A}\right)/T_{\text C}$ is plotted in Fig.~\ref{ris:fig5}(d) as a function of the relative change in the propagation vector $k_{\text s}(x)/k_{\text s}(0)$ for all the compounds ($k_{\text s}(0)$ is the propagation vector of pure Cu$_2$OSeO$_3$), a linear scale is obtained for the (Cu,Ni)$_2$OSeO$_3$ series. On the contrary, (Cu,Zn)$_2$OSeO$_3$ samples do not display a correlation between these two parameters and do not line up with the trend demonstrated by (Cu,Ni)$_2$OSeO$_3$. In order to construct a uniform scaling, another microscopic parameter should be considered.

The magnetic structure of cubic helimagnets with $P2_13$ symmetry has been successfully understood within the Bak-Jensen theory, which considers the free-energy expansion in terms of a slow-varying spin density~\cite{Bak}. Such an expansion is valid if the magnetic structure has a long modulation (spiral period is much greater than the lattice constant). According to the Bak-Jensen model~\cite{Bak,Maleyev,Grigoriev4}, $k_{\text s}$ is defined by the ratio $k_{\text s} = D/J$, where $D$ --- the DMI constant and $J$ is the isotropic Heisenberg exchange constant. Since $T_{\text C} \propto J$, the change in $D$ can be estimated by the known changes in $T_{\text C}$ and $k_{\text s}$. When the nonmagnetic Zn impurities are introduced, the average magnitudes of the exchange and DMI constant $\langle J\rangle$, $\langle D\rangle$ (where $\langle ...\rangle$ denotes the average over nearest-neighbor sites) change in accordance. Thus, the resulting value of $k_{\text s}$ remains almost unchanged. When Cu is replaced by Ni, the Ni site might contribute a different $D/J$ ratio. The $\sim 20$\% increase in $k_{\text s}$ for $x(\text{Ni}) = 0.04$ cannot be accounted for solely by $\sim 5$\% reduction in $T_{\text C}$ (reduction in $\langle J\rangle$). Therefore, one can conclude that the DMI is increased by $\sim 15$\% when Cu is replaced by Ni in Cu$_2$OSeO$_3$.

The relative change in DMI constant calculated within the Bak-Jensen model is shown in Fig.~\ref{ris:fig5}(e). This plot demonstrates that, indeed, the DMI is reduced when the magnetic sublattice is diluted by the nonmagnetic ions, whereas Ni ions bring to the system a larger $D$. The enhancement (suppression) of the SkL stability upon Ni (Zn) substitution is now clearly evidenced when the $T$-width of the SkL phase is plotted versus the extracted $D$ constant of each compound~[Fig~\ref{ris:fig5}(f)]. The uniform scaling can be applied to the whole dataset. As can be seen, the SkL stability is linearly proportional to the DMI constant. The extrapolation of the obtained fit on the lower $D$ values yields the prediction that the SkL phase would be suppressed completely in the system if the DMI is tuned to the value that amounts to $\sim 85$\% of the original constant.

It is worth to note that the enhancement of the SkL stability may not be solely attributed to the changes in the DMI, as the modified cubic anisotropy might also play a role in the skyrmion stabilization~\cite{Chacon18,Qian18}. Utesov \textit{et al.}~\cite{Utesov} showed that the defects in DMI constant in systems with low defect concentrations lead to a long-range corrections to the spiral period. This might explain why no local distortions in the spiral pitch occur due to disorder. Instead, the local defects in DMI homogeneously change the spiral period in the whole system. Further theoretical studies are required to investigate if the local changes in the cubic anisotropy can lead to a similar long-range influence on the magnetic texture.

\section{conclusion}

To conclude, we used small-angle neutron scattering to study the details of the magnetic structure of a chemically-substituted Cu$_2$OSeO$_3$, which is a model compound to study magnetic skyrmions. We utilized samples where mangetic Cu ions are substituted by either nonmagnetic Zn or magnetic Ni within the limit of low impurity concentration. We experimentally determined that all the substituted compounds posses a helimagnetic structure and host the SkL phase in applied magnetic fields at temperatures close to $T_{\text C}$, similarly to the pristine Cu$_2$OSeO$_3$. We observed that Ni-substitution leads to a significant enhancement of the temperature stability of the skyrmions and also results a noticeable expansion of the helical propagation vector (reduction of the skyrmion size). On the contrary, the effect of Zn substitution only causes a moderate reduction of the thermal stability of the skyrmions, retaining the magnitude of the propagation vector. The concentration-dependent change in the skyrmion stability can be explained by the induced change in the strength of the Dzyaloshinskii-Moriya interaction.

\section*{Acknowledgments}

A.S.S. acknowledges support from the International Max Planck Research School for Chemistry and Physics of Quantum Materials (IMPRS-CPQM). The work at the TU Dresden was funded by the German Research Foundation (DFG) in the framework of the Collaborative Research Center SFB 1143 (project C03), W\"{u}rzburg-Dresden Cluster of Excellence \textit{ct.qmat} (EXC 2147, project-id
39085490), and the Priority Program SPP 2137 ``Skyrmionics''.

\end{document}